\documentclass{emulateapj}

\usepackage{color}

\def\simlt{\lower.5ex\hbox{$\; \buildrel < \over \sim \;$}}
\def\simgt{\lower.5ex\hbox{$\; \buildrel > \over \sim \;$}}

\def\gsim{\lower 2pt \hbox{$\, \buildrel {\scriptstyle >}\over
{\scriptstyle \sim}\,$}}
\def\lsim{\lower 2pt \hbox{$\, \buildrel {\scriptstyle <}\over
{\scriptstyle \sim}\,$}}

\def\deg{\ifmmode ^{\circ}
         \else $^{\circ}$\fi}
\def\pdeg{\ifmmode
           $\setbox0=\hbox{$^{\circ}$}\rlap{\hskip.11\wd0 .}$^{\circ}
     \else \setbox0=\hbox{$^{\circ}$}\rlap{\hskip.11\wd0 .}$^{\circ}$\fi}

\def\vturb{\ifmmode v_\mathrm{turb}\else $v_\mathrm{turb}$\fi}
\def\tcut{\ifmmode T_\mathrm{cut}\else $T_\mathrm{cut}$\fi}
\def\rfeii{\ifmmode R_\mathrm{Fe~II}\else $R_\mathrm{Fe~II}$\fi}
\def\aox{\ifmmode \alpha_\mathrm{ox}\else $\alpha_\mathrm{ox}$\fi}
\def\ax{\ifmmode \alpha_\mathrm{x}\else $\alpha_\mathrm{x}$\fi}
\def\aix{\ifmmode \alpha_\mathrm{ix}\else $\alpha_\mathrm{ix}$\fi}
\def\auv{\ifmmode \alpha_\mathrm{uv}\else $\alpha_\mathrm{uv}$\fi}

\def\alphaox{$\alpha_{\mathrm{ox}}$}
\newcommand{\siiii}{Si~{\sc iii]}}
\newcommand{\siiv}{Si~{\sc iv}}
\newcommand{\ciii}{{\sc C~iii]}}
\newcommand{\civ}{{\sc C~iv}}
     
\def\pc{\ifmmode \mathrm{pc} \else $\mathrm{pc}$ \fi}
\def\mpc{\ifmmode \mathrm{Mpc} \else $\mathrm{Mpc}$\fi}
\def\mpcthree{\ifmmode \mathrm{Mpc}^{-3} \else $\mathrm{Mpc}^{-3}$\fi}
\def\gpcthree{\ifmmode \mathrm{Gpc}^{-3} \else $\mathrm{Gpc}^{-3}$\fi}

\def\kelvin{\ifmmode \mathrm{K} \else {$\mathrm{K}$}\fi}
\def\kev{\ifmmode \mathrm{keV} \else $\mathrm{keV}$ \fi}

\def\lsun{\ifmmode {L_\odot} \else $L_\odot$\fi}
\def\msun{\ifmmode M_\odot \else $M_\odot$\fi}
\def\msunyr{\ifmmode M_\odot~\mathrm{yr}^{-1} \else $M_\odot~\mathrm{yr}^{-1}$\fi}

\def\cosi{\ifmmode {\cos\,i} \else $\cos\,i$\fi}

\def\hii{\ifmmode {\rm H{\sc ii}} \else H~{\sc ii}\fi}
\def\heii{\ifmmode {\rm He{\sc ii}} \else He~{\sc ii}\fi}
\def\mgii{\ifmmode {\rm Mg{\sc ii}} \else Mg~{\sc ii}\fi}
\def\sithree{\ifmmode {\rm Si{\sc iii}} \else Si~{\sc iii}\fi}
\def\sifour{\ifmmode {\rm Si{\sc iv}} \else Si~{\sc iv}\fi}
\def\caii{\ifmmode {\rm Ca{\sc ii}} \else Ca~{\sc ii}\fi}
\def\ciii{\ifmmode {\rm C{\sc iii}]} \else C~{\sc iii}]\fi}
\def\civ{\ifmmode {\rm C{\sc iv}} \else C~{\sc iv}\fi}
\def\mgii{\ifmmode {\rm Mg~{\sc ii}} \else Mg~{\sc ii}\fi}
\newcommand{\oi}{{\sc [O~i]}}
\newcommand{\oii}{{\sc [O~ii]}}
\newcommand{\oiii}{{\sc [O~iii]}}

\newcommand{\nii}{{\sc [N~ii]}}
\newcommand{\neiii}{{[Ne~{\sc iii}]}}
\newcommand{\nev}{{[Ne~{\sc v}]}}
\newcommand{\sii}{{\sc [S~ii]}}

\newcommand{\feii}{Fe~{\sc ii}}
\newcommand{\fevii}{[Fe~{\sc vii}]}

\def\teff{\ifmmode {T_{\rm eff}} \else $T_{\rm eff}$\fi}
\def\tmax{\ifmmode {T_{\rm max}} \else $T_{\rm max}$\fi}

\def\mbh{\ifmmode {M_{\rm BH}} \else $M_{\rm BH}$\fi}
\def\led{\ifmmode L_{\mathrm{Ed}} \else $L_{\mathrm{Ed}}$\fi}
\def\lbolflare{\ifmmode L_{\mathrm{bol,flare}} \else $L_{\mathrm{bol,flare}}$\fi}
\def\lagn{\ifmmode L_{\mathrm{agn}} \else $L_{\mathrm{agn}}$\fi}
\def\lbolagn{\ifmmode L_{\mathrm{bol,agn}} \else $L_{\mathrm{bol,agn}}$\fi}
\def\lbol{\ifmmode L_{\mathrm{bol}} \else $L_{\mathrm{bol}}$\fi}
\def\mdot{\ifmmode {\dot M} \else $\dot M$\fi}
\def\mdoto{\ifmmode {\dot{M}_0} \else  $\dot{M}_0$\fi}
\def\mdotf{\ifmmode {\dot{M}_\mathrm{flare}} \else  $\dot{M}_\mathrm{flare}$\fi}

\def\hnot{\ifmmode H_0 \else H$_0$ \fi}

\def\vkep{\ifmmode v_\mathrm{Kep} \else $v_\mathrm{Kep}$ \fi}
\def\vc{\ifmmode v_\mathrm{c} \else $v_\mathrm{c}$ \fi}

\def\vthree{\ifmmode v_{1000} \else $v_{1000}$ \fi}
\def\vrel{\ifmmode v_\mathrm{rel} \else $v_\mathrm{rel}$ \fi}
\def\vkick{\ifmmode v_\mathrm{kick} \else $v_\mathrm{kick}$ \fi}
\def\vkickz{\ifmmode v_{\mathrm{kick},z} \else $v_{\mathrm{kick},z} $ \fi}
\def\vkicky{\ifmmode v_{\mathrm{kick},y} \else $v_{\mathrm{kick},y} $ \fi}
\def\vchar{\ifmmode v_\mathrm{char} \else $v_\mathrm{char}$ \fi}
\def\eflare{\ifmmode E_\mathrm{flare} \else $E_\mathrm{flare}$ \fi}
\def\ekick{\ifmmode E_\mathrm{kick} \else $E_\mathrm{kick}$ \fi}
\def\ecoll{\ifmmode E_\mathrm{coll} \else $E_\mathrm{coll}$ \fi}
\def\ezero{\ifmmode E_\mathrm{0} \else $E_\mathrm{0}$ \fi}
\def\efac{\ifmmode \xi_\mathrm{E} \else $\xi_\mathrm{E}$ \fi}
\def\tqso{\ifmmode t_\mathrm{QSO} \else $t_\mathrm{QSO}$ \fi}
\def\tflare{\ifmmode t_\mathrm{flare} \else $t_\mathrm{flare}$ \fi}
\def\tzero{\ifmmode t_\mathrm{0} \else $t_\mathrm{0}$ \fi}
\def\tfac{\ifmmode \xi_\mathrm{t} \else $\xi_\mathrm{t}$ \fi}
\def\gfac{\ifmmode f_\mathrm{g} \else $f_\mathrm{g}$ \fi}
\def\lflare{\ifmmode L_\mathrm{flare} \else $L_\mathrm{flare}$ \fi}
\def\fflare{\ifmmode F_\mathrm{flare} \else $F_\mathrm{flare}$ \fi}
\def\nflare{\ifmmode N_\mathrm{flare} \else $N_\mathrm{flare}$ \fi}
\def\tshock{\ifmmode T_\mathrm{shock} \else $T_\mathrm{shock}$ \fi}
\def\rmin{\ifmmode R_\mathrm{1} \else $R_\mathrm{1}$ \fi}
\def\rmax{\ifmmode R_\mathrm{2} \else $R_\mathrm{2}$ \fi}
\def\rbound{\ifmmode R_\mathrm{b} \else $R_\mathrm{b}$ \fi}
\def\pbound{\ifmmode P_\mathrm{b} \else $P_\mathrm{b}$ \fi}
\def\mbound{\ifmmode M_\mathrm{b} \else $M_\mathrm{b}$ \fi}
\def\mbo{\ifmmode M_{\mathrm{b}0} \else $M_{\mathrm{b}0} $ \fi}
\def\ebo{\ifmmode E_{\mathrm{b}0} \else $E_{\mathrm{b}0} $ \fi}
\def\efinal{\ifmmode E_\mathrm{final} \else $E_\mathrm{final} $ \fi}
\def\tbound{\ifmmode t_\mathrm{b} \else $t_\mathrm{b}$ \fi}
\def\tagn{\ifmmode t_\mathrm{AGN} \else $t_\mathrm{AGN}$ \fi}
\def\rlim{\ifmmode R_\mathrm{lim} \else $R_\mathrm{lim}$ \fi}
\def\vlim{\ifmmode v_\mathrm{lim} \else $v_\mathrm{lim}$ \fi}
\def\vphi{\ifmmode v_\phi \else $v_\phi$ \fi}
\def\mlim{\ifmmode M_\mathrm{lim} \else $M_\mathrm{lim}$ \fi}
\def\tlim{\ifmmode t_\mathrm{lim} \else $t_\mathrm{lim}$ \fi}
\def\llim{\ifmmode L_\mathrm{lim} \else $L_\mathrm{lim}$ \fi}
\def\fqso{\ifmmode f_\mathrm{QSO} \else $f_\mathrm{QSO}$ \fi}

\def\hbeta{\ifmmode \rm{H}\beta \else H$\beta$\fi}
\def\hbetan{\ifmmode \rm{H}\beta_{\rm n} \else H$\beta_{\rm n}$\fi}
\def\hgamma{\ifmmode \rm{H}\gamma \else H$\gamma$\fi}
\def\hdelta{\ifmmode \rm{H}\delta \else H$\delta$\fi}
\def\hepsilon{\ifmmode \rm{H}\epsilon \else H$\epsilon$\fi}
\def\hzeta{\ifmmode \rm{H}\zeta \else H$\zeta$\fi}
\def\halpha{\ifmmode \rm{H}\alpha \else H$\alpha$\fi}
\def\lalpha{\ifmmode \rm{Ly}\alpha \else Ly$\alpha$}

\def\dvhb{\ifmmode \Delta v_{\hbeta} \else $\Delta v_{\hbeta}$\fi}
\def\dvmg{\ifmmode \Delta v_{\rm{Mg}} \else $\Delta v_{\rm{Mg}}$\fi}

\def\muobs{\ifmmode {\mu_{o}} \else  $\mu_{o}$ \fi}
\def\cosi{\ifmmode {\mathrm{cos}\,i} \else $\mathrm{cos}\,i$\fi}

\def\teff{\ifmmode {T_{eff}} \else $T_{eff}$ \fi}
\def\tmax{\ifmmode {T_{max}} \else $T_{max}$ \fi}

\def\tauh{\ifmmode {\tau_{\rm H}} \else $\tau_{\rm H}$ \fi}

\def\yr{\ifmmode {\rm yr} \else  yr \fi}
\def\kms{\ifmmode \rm km~s^{-1}\else $\rm km~s^{-1}$\fi}
\def\cm{\ifmmode {\rm cm} \else  cm \fi}
\def\cmmitwo{\ifmmode \rm cm^{-2} \else $\rm cm^{-2}$\fi}
\def\cmmithree{\ifmmode \rm cm^{-3} \else $\rm cm^{-3}$\fi}
\def\cmtwos{\ifmmode \rm cm^{-2}~s^{-1}\else $\rm cm^{-2}~s^{-1}$\fi}
\def\cmps{\ifmmode \rm cm~s^{-1}\else $\rm cm~s^{-1}$\fi}
\def\cmpsps{\ifmmode \rm cm~s^{-2}\else $\rm cm~s^{-2}$\fi}
\def\kmps{\ifmmode \rm km~s^{-1}\else $\rm km~s^{-1}$\fi}
\def\kmpspmpc{\ifmmode \rm km~s^{-1}~Mpc^{-1} \else
    $\rm km~s^{-1}~Mpc^{-1}$\fi}
  
\def\gcmthree{\ifmmode \rm g~cm^{-3} \else $\rm g~cm^{-3}$\fi}
\def\gcmtwo{\ifmmode \rm g~cm^{-2} \else $\rm g~cm^{-2}$\fi}
   
\def\erg{\ifmmode {\rm erg} \else $\rm erg$ \fi}
\def\ergps{\ifmmode {\rm erg~s^{-1}} \else $\rm erg~s^{-1}$ \fi}
\def\ergcms{\ifmmode \rm erg~cm^{-2}~s^{-1} \else $\rm erg~cm^{-2}~s^{-1}$ \fi}
\def\ergcmshz{\ifmmode \rm erg~s^{-1}~cm^{-2}~Hz^{-1} \else $\rm
erg~cm^{-2}~s^{-1}~Hz^{-1}$ \fi}
\def\ergcmsa{\ifmmode \rm erg~cm^{-2}~s^{-1}~\AA^{-1} \else $\rm
erg~cm^{-2}~s^{-1}~\AA^{-1}$ \fi}
\def\ergshz{\ifmmode \rm erg s^{-1} Hz^{-1} \else
   $\rm erg s^{-1} Hz^{-1}$ \fi}

\def\lam{\ifmmode {\lambda} \else {$\lambda$} \fi}
\def\llam{\ifmmode {L_\lambda} \else  $L_\lambda$ \fi}
\def\lamLlam{\ifmmode \lambda L_{\lambda}(5100) \else {$\lambda L_{\lambda}(5100)$} \fi}
\def\nuLnu{\ifmmode \nu L_{\nu}(5100) \else {$\nu L_{\nu}(5100)$} \fi}
\def\ilam{\ifmmode {I_\lambda} \else  $I_\lambda$ \fi}
\def\flam{\ifmmode {F_\lambda} \else  $F_\lambda$ \fi}
\def\inu{\ifmmode {I_\nu} \else  $I_\nu$ \fi}
\def\fnu{\ifmmode {F_\nu} \else  $F_\nu$ \fi}
\def\bnu{\ifmmode {B_\nu} \else  $B_\nu$ \fi}

\def\msigma{\ifmmode M_{\sigma} \else $M_{\sigma}$\fi}
\def\mbulge{\ifmmode M_{\mathrm{bulge}} \else $M_{\mathrm{bulge}}$\fi}
\def\mgal{\ifmmode M_{\mathrm{gal}} \else $M_{\mathrm{gal}}$\fi}
\def\lgal{\ifmmode L_{\mathrm{gal}} \else $L_{\mathrm{gal}}$\fi}
\def\lbulge{\ifmmode L_{\mathrm{bulge}} \else $L_{\mathrm{bulge}}$\fi}
\def\mgalstar{\ifmmode M^*_{\mathrm{gal}} \else $M^*_{\mathrm{gal}}$\fi}

\def\mbhsigstar{\ifmmode M_{\mathrm{BH}} - \sigma_* \else $M_{\mathrm{BH}} - \sigma_*$ \fi}
\def\deltalogmbh{\ifmmode \Delta~{\mathrm{log}}~M_{\mathrm{BH}} \else $\Delta$~log~$M_{\mathrm{BH}}$\fi}

\def\sigstar{\ifmmode \sigma_* \else $\sigma_*$\fi}
\def\sigthree{\ifmmode \sigma_{\mathrm{[O~III]}} \else $\sigma_{\mathrm{[O~III]}}$\fi}
\def\sigtwo{\ifmmode \sigma_{\mathrm{[O~II]}} \else $\sigma_{\mathrm{[O~II]}}$\fi}
\def\signl{\ifmmode \sigma_{\mathrm{NL}} \else $\sigma_{\mathrm{NL}}$\fi}
\def\wthree{\ifmmode {\rm FWHM({[O~III]})} \else $FWHM({[O~III]})$ \fi}
\def\wtwo{\ifmmode {\rm FWHM({[O~II]})} \else $FWHM({[O~II]})$ \fi}
\def\mthree{\ifmmode M_{\mathrm [O~III]} \else $M_{\mathrm [O~III]}$ \fi}
\def\mtwo{\ifmmode M_{\mathrm [O II]} \else $M_{\mathrm [O II]}$ \fi}

\def\lbreak{\ifmmode L_{\mathrm{break}} \else $L_{\mathrm{break}}$\fi}
\def\lcut{\ifmmode L_{\mathrm{cut}} \else $L_{\mathrm{cut}}$\fi}

\received{2010 June 9}
\accepted{2010 August 5}

\slugcomment{ApJ in press}

\shortauthors{Shields, Ludwig, \& Salviander}
\shorttitle{\feii\  Emission in AGN}

\begin{document}

\title{\feii\  Emission in AGN: The Role of Total and Gas-Phase Iron Abundance}

\author{  Gregory~A. Shields\altaffilmark{1}, Randi R. Ludwig\altaffilmark{1}, Sarah~Salviander \altaffilmark{1,2}}

\altaffiltext{1}{Department of Astronomy, University of Texas, Austin,
TX 78712; shields@astro.as.utexas.edu; randi@astro.as.utexas.edu; triples@astro.as.utexas.edu} 

\altaffiltext{2}{Department of Physics, Southwestern University, Georgetown, TX 78626}

\begin{abstract}

Active galactic nuclei (AGN) have \feii\ emission from the broad line region (BLR) that differs greatly in strength from object to object. We examine the role of the total and gas-phase iron abundance in determining \feii\ strength.   Using AGN spectra from the Sloan Digital Sky Survey (SDSS) in the redshift range of $0.2Ê< zÊ<Ê0.35$,  we measure the Fe/Ne abundance of the narrow line region (NLR) using the \fevii/\nev\ line intensity ratio. We find no significant difference in the abundance of Fe relative to Ne in the NLR as a function of \feii/\hbeta.   However, the \nii/\sii\ ratio increases a by a factor of 2 with increasing \feii\ strength.  This indicates a trend in N/S abundance ratio, and by implication in the overall metallicity of the NLR gas, with increasing \feii\ strength. 

We propose that the wide range of \feii\ strength in AGN largely results from the selective depletion of Fe into grains in the low ionization portion of the BLR.  Photoionization models show that the  strength of the optical \feii\ lines varies almost linearly 
with gas-phase Fe abundance, while the ultraviolet \feii\ strength varies more weakly.  Interstellar depletions of Fe can be as large as two orders of magnitude, sufficient to explain the wide range of optical \feii\ strength in AGN.  This picture is consistent with the similarity of the BLR radius to the dust sublimation radius and with indications of \feii\ emitting gas flowing inwards from the dusty torus.

\end{abstract}


\section{Introduction}
\label{sec:intro}

The broad emission-line spectrum of quasars often includes strong \feii\ in the optical and ultraviolet.  The difference  between the weakest and strongest optical \feii\ emission exceeds a factor of 10,
measured as equivalent width (EW) or as \feii/\hbeta\ line ratio.  For recent discussions of \feii\ intensities in AGN, and references to earlier work,
see \citet{kovacevic10} and \citet{ferland09}.
The \feii\ strength anti-correlates with the strength of the narrow \oiii\ emission line.  This
trend along with several associated correlations  defines the  so-called ``Eigenvector~1''  (EV1), which
 characterizes some of the most conspicuous differences among the properties of AGN
\citep[hereinafter BG92]{boroson92}.   The quest for physical drivers of EV1 has inspired a number of
studies \citep[e.g.,][]{wills99, marziani03}.   \citet{boroson02} and \citet{netzer07} find that
\feii\ increases with Eddington ratio $L/\led$, as originally suggested by BG92.  However, the physics underlying this correlation remains unclear.    There is even debate as to whether the \feii\ emission is entirely powered by the ionizing continuum of the central source, or comes in some measure from a mechanically heated region \citep{wills85, collin00, sigut03, bruhweiler08}.
 In this situation, any observational clues to the nature of EV1 and the great range of \feii\ strength are of value.   

It is generally assumed that the strength of Fe II emission is driven by physical conditions within the BLR,
such as ionizing continuum, BLR density and geometry, column density, and turbulent velocity.   However, a high abundance of Fe has been discussed to help produce the strongest \feii\ observed \citep{wills85, collin88, hamann93}.   The utility of \feii\ to assess the Fe abundance in high redshift QSOs has received considerable interest, in the context of galactic chemical evolution \citep[and references therein]{hamann93, verner03, baldwin04, netzer07}.   Here we assess the importance of the abundance of Fe relative to the $\alpha$-elements, the overall metallicity  of the nuclear gas,  and the depletion of Fe into grains for the strength of \feii\ emission in QSOs.

\section{QSO Sample and Measurements}
\label{sec:meas}

\begin{figure}[ht]
\begin{center}
\plotone{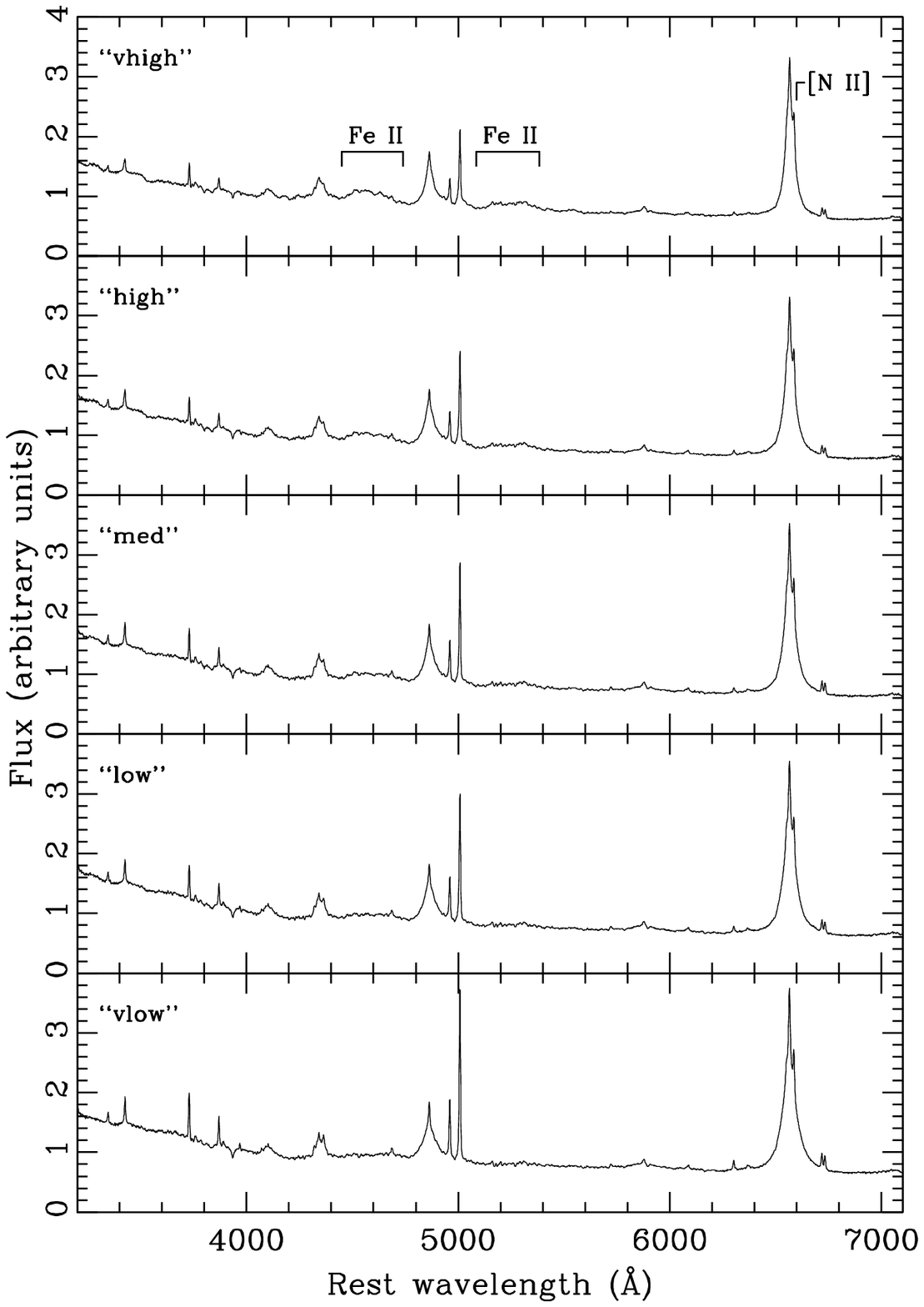}
\figcaption[]{
Composite SDSS spectra for the five bins in optical \feii\ strength.  Vertical axis gives
specific flux \flam.  Note the broad \feii\ blends at $\lambda 4570$ and $\lambda5250$. See text for discussion.
\label{fig:comp} 
}
\end{center}
\end{figure}

\begin{figure}[ht]
\begin{center}
\plotone{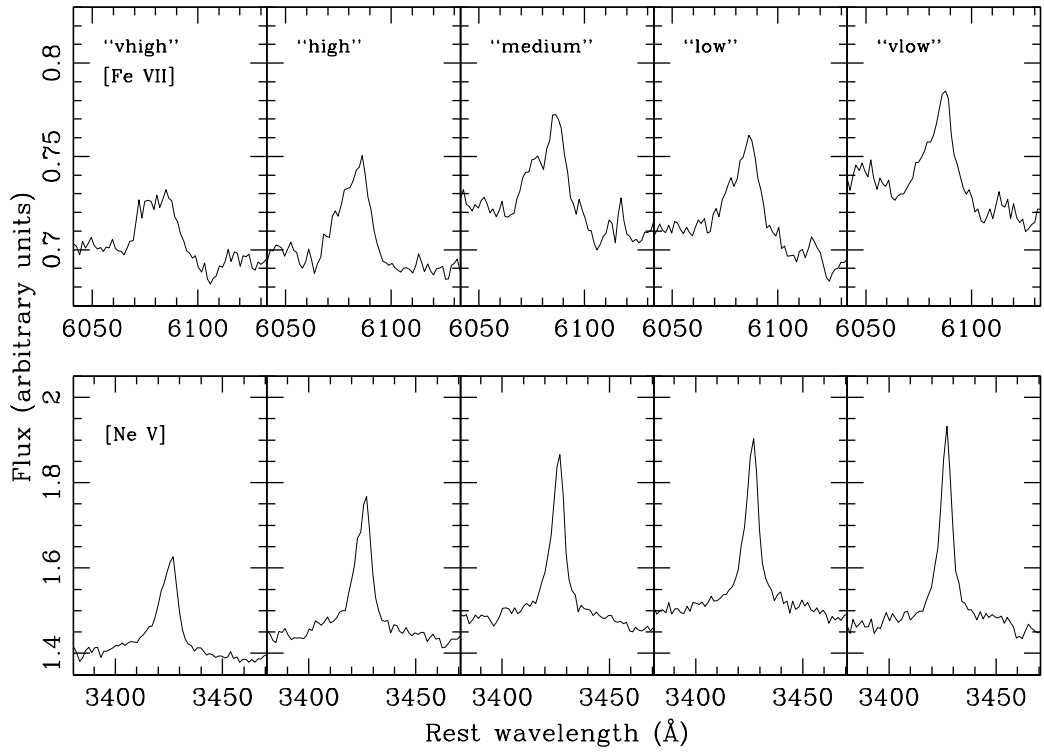}
\figcaption[]{
\fevii\ and \nev\ narrow emission lines in the composite spectra.  See text for discussion.
\label{fig:fene} 
}
\end{center}
\end{figure}

	We investigated the influence of differing abundances in quasars on their optical \feii\ emission strength by studying the optical emission-line properties of a sample of QSOs from the Sloan Digital Sky Survey (SDSS) \footnote{The SDSS website is
http://www.sdss.org.}.   It is difficult to measure abundances within the BLR directly, 
because it is a region of high density, line width, and line optical depth. 
On the assumption that the abundance in the NLR and BLR is the same for a given object (see below), we used the narrow emission-line spectrum to assess abundances in the NLR.  We considered two key line ratios:
(1)  The  \fevii/\nev\ intensity ratio gives a measure the Fe/Ne abundance ratio.  (2) The
\nii/\sii\ ratio gives a measure  of the N/S abundance ratio, which is in turn a secondary indicator of
the overall metallicity of the gas.
	Our sample consists of 1571 quasars from SDSS Data Release 7 (DR7).  These objects were selected in the manner of the ``HO3'' sample of \citet{salviander07}, with the additional requirement of a signal-to-noise (S/N) ratio greater than 10 in the continuum at $\lambda5100$ rest wavelength.   The flux and equivalent width (EW) of the broad \hbeta\ line and the optical \feii\ blends were measured with the aid of a spectrum fitting program described by \citet{salviander07}, using a template fitting procedure to establish the flux in \feii\ relative to the local continuum.  We characterized the \feii\ emission strength using the flux ratio of the \feii~4570~\AA\ blend to broad \hbeta, following BG92.    We used quasars at redshifts $0.2 < z < 0.35$ to ensure coverage of both \nev~$\lambda3425$ and \fevii~$\lambda6087$.   In order to bring out the weak \fevii\ line, we binned the 1571 objects by \feii\  strength,  and made five composite spectra of  ``very weak,''  ``weak,''  ``medium,''  ``strong,'' and ``very strong''  \feii\ emission.  These composites had 312 to 315 objects, within bins bounded by \feii/\hbeta\ values of 0, 0.215, 0.372, 0.524, 0.708, and 1.82, respectively.   Individual spectra were corrected for Galactic reddening using the extinction values $A_g$ given by the SDSS pipeline, and normalized to a mean flux density \flam\ of unity using all wavelength points in a particular spectrum. The individual spectra in each group were shifted in wavelength to the rest frame and re-binned to a common wavelength grid at a spacing of 1.41~\AA.  The adopted composite spectrum was a mean of the rebinned \flam\ for all contributing spectra at a given wavelength.  The composite spectra are shown in Figures \ref{fig:comp} and \ref{fig:fene}.  The region of the \fevii\ and \nev\ line is shown in Figure \ref{fig:fene}.  For a discussion of issues involving composite spectra of QSOs, see \citet{vandenberk01}, and references therein.
	
\begin{figure}[ht]
\begin{center}
\plotone{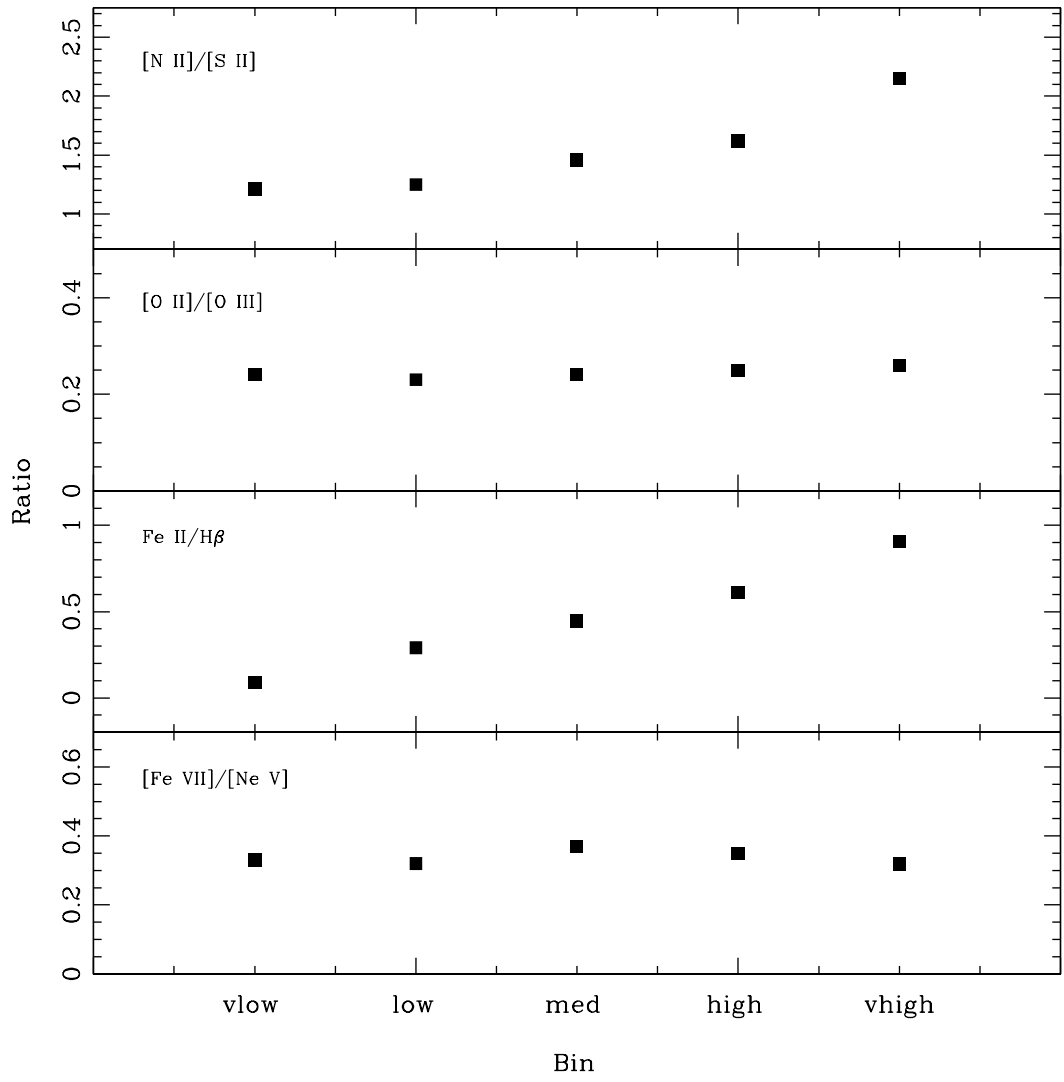}
\figcaption[]{
Narrow emission-line intensity ratios ratios for the five composite spectra.
See text for discussion including errors.
\label{fig:trend} 
}
\end{center}
\end{figure}

From these composites, we measured the emission-line fluxes of a number of lines,
including \oiii~$\lambda5007$, \oii~$\lambda3727$, \neiii~$\lambda3869$, \nev~$\lambda3425$,
\sii~$\lambda\lambda6716, 6730$, and \fevii~$\lambda6087$.   The results are given in Table~1 and
Figure~\ref{fig:trend}, where the \feii/\hbeta\ values are averages of the values for the individual spectra that
compose each composite.   Most lines were measured using a Gaussian fit
with the IRAF task SPLOT  
\footnote{IRAF is distributed by the National Optical Astronomy Observatories, which are operated by the Association of Universities for Research in Astronomy, Inc., under cooperative agreement with the National Science Foundation.}.  
The broad Balmer emission lines (\halpha, \hbeta) were
measured using a Lorentzian profile.  Estimated uncertainties are $10\%$ for
the stronger lines, including continuum placement and faithfulness of the fit.  For
\fevii\ and \nev, the uncertainty is as much as $20\%$, based on noise, continuum uncertainty, and
 the presence of a strong blue wing on both lines that we excluded from the fit.  
The \nii~$\lambda6583$ and $\lambda6548$
lines were fairly weak bumps on the wings of the broad \halpha\ line and relatively difficult to
measure.  Therefore we measured the \nii\  intensity by subtracting from the \halpha\  -- \nii\ blend
a doublet with the theoretical 3.0-to-1 intensity ratio, each line having a Gaussian
profile with a central wavelength and width based on the redshift and line width of 
\sii~$\lambda6716$.  The intensity of the doublet was adjusted so that the \halpha\ line
had a smooth profile with no visible residual intensity or over-subtraction of \nii.  Error
bars were estimated by determining \nii\ intensities giving a slight under- or
over-subtraction as judged by eye.   This gave an uncertainty of about $\pm12\%$ for each composite.

\begin{deluxetable}{lccccc} 
\tablewidth{0pt} 
\tablecaption{Emission Line Ratios for Composite Spectra \label{t:bins}} 
\tablehead{ 
\colhead{Emission} & 
\colhead{} & 
\colhead{} & 
\colhead{Intensity Ratio} & 
\colhead{} & 
\colhead{} \\ 
\colhead{Lines}   & 
\colhead{vlow}  & 
\colhead{low}   & 
\colhead{med}   & 
\colhead{high}  & 
\colhead{vhigh}} 
\startdata 
\nii/\sii & 1.21 & 1.25 & 1.46 & 1.62 & 2.15 \\ 
\nii/\oii & 0.96 & 1.25 & 1.22 & 1.55 & 2.29 \\ 
\oii/\oiii & 0.24 & 0.23 & 0.24 & 0.25 & 0.26 \\ 
\sii/\oii & 0.80 & 0.95 & 0.83 & 0.95 & 1.07 \\ 
\fevii/\nev & 0.33 & 0.32 & 0.37 & 0.35 & 0.32 \\ 
\nev/\neiii & 0.97 & 1.10 & 1.15 & 1.61 & 1.81 \\ 
\halpha/\hbeta & 3.64 & 3.39 & 4.14 & 3.52 & 3.04 \\ 
\sii (6720/4072) & 9.2: & 10.3: & 9.7: & 9.9: & 13.5: \\ 
\sii (6716/6731) & 1.19 & 1.05 & 1.07 & 1.10 & 1.08 \\ 
\feii/\hbeta  & 0.11 & 0.29 & 0.44 & 0.61 & 0.91 \\ 
\enddata 
 
\tablecomments{Intensity ratio for emission lines measured from composite 
spectra binned by broad \feii\ strength. Values refer to the narrow 
emission lines except for the Balmer lines and \feii.  Intensities include both lines for the
\oii\ and \sii\ doublets but only the stronger line for \nii, \oiii, \neiii, and \nev.  Colon indicates large 
uncertainty.   
See text for discussion.} 
 
\end{deluxetable}

Our results will be discussed in terms of trends of observed line ratios.  For actual ionic abundances,  collision strengths from Berrington et al. (2000) and Osterbrock \& Ferland (2006) lead to the relation
\begin{equation}
n(Fe^{+6})/n(Ne^{+4}) = 0.91\, I(\lambda6087)/I(\lambda3425)
\end{equation}	
for an assumed T=15,000 K based on photoionization models.  Likewise, for the \nii/\sii\ ratio we have
\begin{equation}
n(N^+)/n(S^+) = 6.0 \, I(\lambda\lambda6584, 6548)/I(\lambda\lambda6716, 6730)
\end{equation}	
for an assumed T=12,000 K.  These expressions give the ionic abundance ratios, if collisional de-excitation is unimportant.

\section{Results for the Narrow Line Region}
\label{sec:results}

\subsection{Iron}
\label{subsec:iron}

\citet{nussbaumer70} suggested that \fevii/\nev\ should be a good measure of the Fe/Ne ratio, based on the similarity
of the ionization potentials.   There is no significant  trend in \fevii/\nev\ in our composite spectra.
The  \feii/\hbeta\ ratio varies by a factor of 8 from the ``very low'' to  ``very high''  composite.  Taking neon to represent the $\alpha$-elements in general, we conclude that  differences in Fe abundance, relative to the $\alpha$-elements is not a significant cause of the observed range of \feii\ emission strength, from a statistical point of view.
In particular, overabundances of Fe/O and Fe/Mg of a factor of 2 to 10, as motivated by chemical evolution models and attempts to fit
the \feii/\mgii\ ratio with photoionization models \citep{wills85, hamann93}, appear to be ruled out.

This conclusion is based on the assumption that the \fevii/\nev\ ratio is a faithful measure of Fe/Ne, and that Fe is not significantly
depleted into grains in the \fevii\ zone of the NLR.    \citet{ferguson97a} and \citet{nagao03}
 conclude that refractory elements are not depleted in the coronal
line region of the NLR, but they also find that \fevii\ and \nev\  do not come from the same
place in ``locally optimally emitting cloud'' (LOC) models of the NLR.   
Here we assume that any ionization correction  for  Fe$^{+6}$/Ne$^{+4}$ in the NLR
does not change systematically with the broad line \feii\ intensity among our composites.    We further assume that the
abundances in the BLR are similar to those in the NLR.  However, it is possible that intense star formation in the nucleus
may give chemical enrichment on a spatial scale smaller than the NLR \citep[see][and references therein]{hamann07, hamann99}.

\subsection{Ionization}
\label{subsec:ion}

The composite spectra were originally constructed to assess the \fevii\ strength in the NLR, but they also afford an opportunity to examine other narrow
line ratios for any systematic dependence on \feii\ strength.  One issue is the level of ionization of the gas. 
Table 1 and Figure \ref{fig:trend} show that the \oii/\oiii\ ratio is closely similar for the various \feii\ bins.  There is a significant increase
in \nev/\neiii\ with increasing  \feii\ strength, reflecting a $\sim25$\%\  {\em decrease} in the \nev\ equivalent width across
the bins and a much larger decrease in the EW of \neiii.  
The narrow \heii~$\lambda4686$ line to \oiii~$\lambda5007$ intensity ratio (not given in Table 1) 
shows a similar, though less continuous, increase from 0.06 to 0.13 across the five bins; 
this reflects a substantially constant equivalent width of narrow \heii\ together with
a systematic decrease in the EW of $\lambda5007$ from 25 to 12~\AA\ across the bins.  
The anticorrelation of \feii\ and \oiii\ is well known (BG92).   Although the constancy of \fevii/\nev\ could result from offsetting ionization
and abundance trends, a straightforward interpretation is that the relative size of the zone containing O$^{+2}$ and Ne$^{+2}$ decreases with
increasing \feii\ without a major effect on the Fe$^{+6}$/Ne$^{+4}$ ionization correction.

\subsection{Reddening}
\label{subsec:red}

The \sii/\oii\ ratio shows an increase of about 30\%\ with increasing \feii\ strength.  This may be an indication of a modest increase
in reddening of the NLR with increasing \feii.   The \sii\ $I(\lambda6720)/I(\lambda4072)$ ratio is quite uncertain but is
consistent with a reddening trend of this magnitude.
This trend is not evident in the broad $I(\halpha)/I(\hbeta)$  intensity ratio.

\subsection{Nitrogen}
\label{subsec:nitrogen}

The \nii/\sii\ ratio shows a systematic increase by a factor of 2 from the very low to the very high \feii\ bins.  This ratio is insensitive
to reddening and electron temperature.  The N$^+$ and S$^+$ ions have similar ionization potentials and occupy similar
zones of the nebular ionization structure.  This and the constancy of \oii/\oiii\ suggests that the trend in \nii/\sii\ is not 
a result of ionization of the NLR.  The trend of increasing \nii\ strength is evident in the \nii/\oii\ and \nii/\oiii\ ratios as well.  These results imply a real trend in the N/S and N/O chemical abundance ratios with increasing \feii\ strength, amounting to a factor of 2 from the ``very low''  to the ``very high'' \feii\ strengths.   Nitrogen is largely a secondary nucleosynthetic product, so that
N/O increases with O/H.  The \hii\ region results of \citet{vanzee98} show N/O increasing almost linearly with O/H above
$12 + \log \mathrm{O/H} = 8.5$.  Assuming that O/H is in this range in the AGN studied here, then the trend of N/O with \feii\
implies an increase of a factor of 2 in O/H.  Although the chemical evolution of AGN host galaxies 
may be complicated \citep{hamann99,hamann02, netzer07},
it may be reasonable to assume that Fe/H varies roughly with O/H for the modest redshifts considered here.  In this case, 
the Fe/H abundance ratio may contribute roughly a factor of 2 to the range in \feii\ strength from our ``very low'' to ``very high'' bins.
This conclusion is consistent with the work of \citet{netzer07}, who find an increase of \feii\ strength with Eddington ratio $L/\led$ in
a large sample of SDSS quasars.  Combining this with indications of an increasing N/C with $L/\led$ in the BLR \citep{scott04, hamann02}, they
argue that the overall metallicity in QSOs increases with \feii\ strength.  However, our results indicate that this makes only a modest contribution to the full range of  optical \feii\ strength in AGN.  Note, however, that our QSO sample is at low redshift, whereas much of the interest in
iron abundances in QSOs has focused on high redshifts.

\section{\feii\ Strength and  X-ray Heating}
\label{xray}

It is widely  assumed that the \feii\ emission is largely powered by the soft X-ray portion of the ionizing spectrum, given that photoionization is indeed the primary excitation mechanism.    The harder photons in the ionizing continuum create an extensive warm, partially ionized zone where \feii\ and other low ions
are subject to collisional excitation and in some cases other excitation mechanisms \citep[see][and references therein]{kwan81, ferland09}.  Does the relative strength of the X-ray continuum drive the range of \feii\ strength observed in AGN?  The following considerations suggest that it does not:

1) A number of  empirical studies have examined correlations of \feii\ strength with the X-ray slope or with the
X-ray/optical ratio \citep[and references therein]{sulentic00}.     Lawrence et al. (1997) studied a sample of AGN with extreme values of  
$\rfeii \equiv I(\lambda4570)/I(\hbeta)$.   The strong \feii\ emitters have X-ray properties, including \alphaox, similar to the weak \feii\ objects.   In particular, the prototype strong \feii\ object, I~Zw~1, has $\aox = -1.4$, an entirely typical value.  Combining their data with the complete sample of \citet{laor97},  Lawrence et al. found little correlation between \alphaox\ and \rfeii, and only an ambiguous correlation with \ax.  
Indeed, Figure 3 of Lawrence et al. 
shows a slight trend in the sense of strong optical \feii\ for {\em weak} X-ray luminosity (steep \aox).  They did find a significant correlation with
the X-ray to IR slope \aix\ in the sense of stronger infrared for stronger \feii. 
Using composite spectra for X-ray bright and X-ray faint QSOs, \citet{green98} found that UV \feii\ was stronger whereas optical \feii\ was weaker for X-ray bright objects.  However, the
differences were small compared to the full range of \feii\ strength among individual AGN.   \citet{leighly07} find a weak X-ray continuum but strong \feii\ emission in
PHL 1811, and discuss other similar examples.   None of these results gives support for the idea that
stronger X-ray continuum drives stronger \feii\ for AGN as a class.   \citet{ferland89} reached a similar conclusion regarding the \caii\ emission
from AGN, which also comes from the partially ionized zone.

2)   In order to explore the expected response of the \feii\ emission to differences in X-ray luminosity (and other parameters)
we have computed a set of models of the BLR using  version 07.02.00 of the photoionization code
Cloudy,  most recently described by \citet{ferland98}.
As a reference model, following \citet{ferland09}, we used  solar abundances, an ionizing flux $\phi = 10^{19}~\cmtwos$, and a gas density $N = 10^{10}~\cmmithree$,  giving an ionization parameter $U \equiv \phi/Nc = 10^{-1.5}$.  The internal turbulent velocity was $u_{\rm turb} = 100~\kms$, and the stopping column density was $10^{23}~\cmmitwo$, in order to include an extensive partially ionized zone.
The  ionizing continuum was the sum of 
(1) a UV component $L_\nu \propto \nu^{-0.5}\mathrm{exp}(-h\nu/k\tcut)$ with $\tcut = 10^{5.7}~\kelvin$  to represent the Big Blue Bump,
and (2) an X-ray power law $L_\nu \propto \nu^{-1}$.  The continuum had a low frequency exponential cutoff below 0.01~Ryd.
 The models used the full treatment of the \feii\ ion (371 levels) and were iterated to convergence of the diffuse radiation field. 
The ratio of the X-ray to the UV component is controlled by the parameter $\aox$.  
For the reference model (Model 1), we used $\aox =  -1.4$, a typical observed value \citep{lawrence97}.
This model has not been optimized to fit a typical AGN emission-line spectrum, but simply serves
as a reference point for exploring the effect on \feii\ of changing various model parameters.   A simultaneous fit to AGN emission-line spectra requires a combination of photoionized clouds with a range of physical conditions \citep{baldwin95}.  

Table~2 gives line intensities from the Cloudy models.  The \halpha, \ciii, \civ, and \mgii\  intensities are reasonable.  \lalpha\ is stronger than observed relative to \hbeta, a familiar problem with photoionization models \citep[e.g.,][]{kwan81}.  The Cloudy output gives \feii\ intensities summed over broad wavelength bands at  1000 -- 2000, 2000 -- 3000, 4000 -- 6000, 6000 -- 7800, and 7800 -- 30000~\AA.  Note that the $\lambda2500$ band includes the broad UV \feii\ bump, and the $\lambda5000$ band contains the prominent optical \feii\ blends at $\lambda4570$ and $\lambda5250$.
Model~1 under-predicts the \feii\ intensity seen in strong \feii\ objects, a common problem with photoionization models as mentioned above.

\begin{deluxetable}{lcccc}
\tablewidth{0pt}
\tablecaption{Photoionization Model Results\label{t:cloudy}}
\tablehead{
\colhead{Model} &
\colhead{1}     &
\colhead{2}     &
\colhead{3}     &
\colhead{4}     \\
\colhead{\alphaox} &
\colhead{-1.4}      &
\colhead{-2.0}      &
\colhead{-1.4}      &
\colhead{-1.4}     \\
\colhead{Fe Depletion} &
\colhead{0.0}   &
\colhead{0.0}   &
\colhead{-1.5}  &
\colhead{-1.5}  \\
\colhead{Other Depletion} &
\colhead{0.0}   &
\colhead{0.0}   &
\colhead{0.0}   &
\colhead{-1.5}}
\startdata
\\
& &$I/I_{\hbeta}$ &  &  \\
\\
1500  & 6.2 & 7.5  & 0.207   & 1.59 \\
2500  & 17.8  & 21.0   & 5.1   & 3.9 \\
5000  & 0.55 & 0.58  & 0.029  & 0.024 \\
7000  & 0.069 & 0.081 & 0.0059 & 0.0045 \\
10000 & 0.42 & 0.44  & 0.030  & 0.024 \\
& & & &  \\
\mgii  & 3.4 & 3.2  & 2.8  & 0.20 \\
\siiii & 1.30& 1.46  & 1.19 & 0.037 \\
\siiv  & 1.56 & 1.75 & 1.42 & 0.104 \\
& & & &  \\
\halpha &  4.6 &  4.6 & 4.5  & 4.6 \\
\lalpha & 28.1 & 32.4 & 25.8 & 21.6 \\
\ciii   &  1.69 &  1.78 & 1.54  & 1.45 \\
\civ    & 13.8 &  13.2 & 12.6  & 11.1 \\
\enddata

\tablecomments{Results of Cloudy photoionization models. Logarithmic depletion relative to hydrogen is given for iron and for other refractory elements. See the text for discussion.}

\end{deluxetable}

In Model 2, we used $\aox =  -2.0$  to explore the effect of weaker X-rays while
remaining within the span of frequently observed values of $\aox$ \citep{lawrence97}.  The \feii\  intensity (relative to \hbeta) actually increased slightly in the X-ray weak model.
We also computed an alternative pair of models with $\phi = 10^{18}~\cmtwos$,  $N = 10^{9}~\cmmithree$, $\tcut = 10^{5.3}~\kelvin$, and no turbulence.  In this case, the \feii\ $\lambda2500$ blend decreased by
0.16~dex and the $\lambda5000$ band by 0.20~dex in going from $\aox = -1.4$ to $-2.0$.    Even in this case, the degree of change is insufficient
to give the full observed range of \feii\ strength.    An extreme systematic variation of \aox\ with \feii\ would be required, for which
the observations give little support.

\section{\feii\ Strength and  Iron Depletion into Grains}
\label{grains}

The above results indicate that changes in the total elemental abundance of Fe and the X-ray luminosity do not cause the wide observed range of \feii\ emission strength among AGN.  We propose instead that \feii\ strength in AGN largely results from differing
degrees of depletion of the gas-phase abundance of iron into grains in the relevant portions of the BLR.  
Gas-phase depletions of refractory elements in the interstellar
medium can be severe.  Iron is depleted by up to 2 orders of magnitude in the interstellar medium and in ionized nebulae (see discussion below).
Such a degree of depletion, if present in AGN with weak \feii\ but not those with strong \feii, can account for much of the observed range of \feii\ strength in AGN.

There has been considerable discussion of refractory element depletions in the NLRs of AGN \citep{ferguson97b}.
 \citet{gaskell81} considered the question of refractory element depletions in the BLR and concluded that depletions of Si, Mg, and Fe as
 severe as those in the ISM did not occur.  This conclusion was based mostly on the \mgii, \sithree, and \sifour\  lines, and the
 intensity of \feii\ in strong \feii\ objects.  Here we suggest that the  degree of depletion of iron varies from object to object, and can be severe
 in objects with weak \feii. 
 In a study of \oi\ and \caii\  emission from AGN,  \citet{matsuoka08} note that depletion of \caii\ might help to reconcile predicted and observed
 intensities and mention the possibility that depletions may affect the
 \feii\ lines.  \citet{ferland89} note that dust mixed with the BLR gas could help to explain low observed values of the \caii\ H and K emission
 lines, relative to the infrared triplet.   As discussed below, the radius of the BLR is interestingly close to the dust sublimation
 radius; and for typical parameters,  grains may exist in the partially ionized zone of the BLR clouds but not in the highly ionized surface layers.
 
The present proposal does not address the long-standing question of how to explain the large \feii\ strength observed in many AGN \citep[and references therein]{wills85, bruhweiler08}.  Rather, it serves to separate the question of \feii\ strength into two parts:  (1) why is \feii\ so strong in some objects, and (2) why is it so weak in others?   We do not attempt to resolve the first question in this paper, except to note that under the present proposal, the \feii\ physical excitation mechanism is freed from the requirement of explaining 
by itself the wide range of observed \feii\ strengths.

\subsection{Photoionization Models}
\label{models}

In order to explore the dependence of \feii\ strength on gas-phase Fe abundance
we have computed two additional Cloudy models (see Table~2).
Model 3 is otherwise identical  to Model 1 but has a depletion
of iron by 1.5~dex.   The $\lambda2500$ feature decreased in intensity by 0.66~dex, and the $\lambda5000$ feature by 1.36~dex,
corresponding to (Fe/H$)^{0.44}$ and (Fe/H)$^{0.91}$, respectively.
In the $N = 10^9~\cmmithree$ model described above, a 1.5~dex depletion of Fe abundance reduced the $\lambda2500$ and $\lambda5000$ features by 0.82~dex  and 1.37~dex, respectively.
Thus, the optical \feii\ intensity varied almost linearly with the gas-phase iron abundance, and the UV \feii\ less strongly. 
These results are consistent with previous photoionization studies.  
\citet{verner03} and \citet{baldwin04}  found  that the \feii\ intensity varies approximately as (Fe/H)$^{0.8}$ for the optical lines and as
(Fe/H)$^{0.4}$ for the UV bump.   Some of these models assume different physical parameters for the BLR from ours,
and the \citet{verner03} models use an 830 level Fe$^+$ model atom as opposed to 371 levels in our models and those of \citet{baldwin04}.  

Other refractory elements may be depleted where iron is.  
In Model 4, the abundances of Al, Mg, Si, Ca, and Fe were all depleted by 1.5~dex.
Table~2 shows that the multiplets \mgii~$\lambda2800$, \sithree~$\lambda1892$, and \sifour~$\lambda1400$  
decreased roughly in proportion to the
gas-phase abundance.  The behavior of the \feii\ multiplets in Model 4 is similar to the case in which only Fe was depleted.

Different refractory elements have different depletions in the ISM,  and different grain compositions may be more or less easily destroyed in the AGN environment.  This may lead to useful diagnostics for grain destruction and gas-phase depletions.
\citet{delgado09} summarize published depletion factors in planetary nebulae:
1/6 to 1/300 for Ca, 1/2 to 1/350 for Al, 1/3 to 1/300 for Fe,  near solar to 1/10 for  Mg, and near solar to 1/20 for Si.
Silicon appears to be depleted by a lesser factor than iron in \hii\ regions.   \citet{garnett95} found Si depletions of only -0.1 to -0.6~dex
in extragalactic H~II regions, significantly less than in dense interstellar clouds.   If Si is more easily restored to the gas
phase in H~II regions, the same may be true in AGN, allowing objects with weak \feii\ to have normal intensities of \sithree\ and \sifour.

\subsection{Geometrical Considerations}
\label{geom}

What might cause widely differing degrees of depletion of iron, and possibly other refractory elements, among AGN?  
From considerations of BLR covering factor, Lyman continuum absorption,
line widths, and reverberation mapping, \citet{gaskell09} argues that the BLR is the inward extension of the dusty torus, with an inflow
velocity a substantial fraction of the orbital velocity.  
Such a picture lends itself to the idea of refractory element depletions in the low ionization zone.  
As material flows inward through the dusty torus, its equilibrium temperature rises as it experiences a stronger radiation field, eventually
reaching the sublimation temperature of  $\sim1500~\kelvin$.   The exact sublimation
radius depends on grain size and composition \citep{laor93}, but there will be a point where the refractory elements are substantially restored to the
gaseous phase.   This sublimation radius is given by  \citet{laor93} as $R_{\mathrm d} = (0.2~\pc) L_{46}$, 
where $L_{46}$ is the AGN luminosity in units of $10^{46}~\ergps$.   If this happens before the material reaches the main region of low ionization line emission, then strong emission in
\feii\ and other lines of refractory elements will occur.  On the other hand, if the dust survives through most of the low ionization zone, then
these emission lines will be weak, because of the lack of emitting ions  and because of attenuation of the
ionizing radiation and line emission by the dust itself.

\citet{netzer93}  suggested that the survival of dust outside the sublimation radius leads to extinction of the ionizing continuum and suppression of line emission, setting a natural limit to radius of the BLR \citep[see also][]{laor07}.  
Infrared variability studies give a radius of the dusty torus just outside the BLR
\citep{suganuma06}, supporting this picture.
 Here we suggest that the relationship between the sublimation radius and the BLR outer boundary differs from object to object.
Critical to this picture is the actual radius of \feii\ emission relative to the dust sublimation radius.
Reverberation studies and line widths indicate that  the lower ionization lines often come from
larger radii in the BLR \citep{peterson97, sulentic00}.  This may result from some combination of actual ionization stratification and
line emissivity effects as considered in the LOC model \citep{baldwin95}.  There have been few successful reverberation studies of the \feii\ 
emitting radius in AGN.    For the Seyfert galaxy NGC 5548, \citet{maoz93} found a similar time lag for the UV \feii\ lines as for \lalpha.  In a reverberation study of Ark~120, which has strong optical \feii, \citet{kuehn08} found that the lag for the optical \feii\ lines was ill-defined but may be around 300~days, larger than for \hbeta.
They estimate a dust sublimation radius in this object of $\sim460$ light days, and conclude that within the uncertainties ``it is plausible that the optical \feii\ emission is produced at or just inside the dust sublimation radius.''   In a study of line profiles of AGN in SDSS, 
\citet{hu08} conclude that the
\feii\ emission comes from an inflowing zone in the outer part of the BLR.  Taken together, these studies are consistent with \feii\ emission from gas entering the BLR from the dusty torus.

A refinement of this picture takes account of the attenuation of the ionizing radiation field with depth within the ionized cloud
or ionized surface layer of the disk or torus.
Reverberation measurements of the BLR radius (often involving the \hbeta\ line) typically give a radius smaller than
the dust sublimation radius by a factor of order 2, based on radii and luminosities given by \citet{bentz09}.  Thus, refractory grains should not survive at the illuminated face of
a BLR cloud.  However, the radiation field diminishes with depth in the cloud, allowing dust to survive at large depths. 
As the dusty gas flows from deep in the torus toward the irradiated surface, the ambient radiation field intensifies
and the grain temperature rises, reaching the sublimation point at some depth in the ionization structure.   Grain
equilibrium temperatures calculated by the Cloudy program support this picture.  In the reference model (Model 1) described
above, the temperature of silicate grains with a radius of 0.094~microns is 680~\kelvin\  at the maximum column depth of
$10^{23}~\cmmitwo$.  This allows grains to survive at this depth, even though the incident flux at the cloud face would easily evaporate refractory grains.  The grain
temperature rises with decreasing depth, reaching the sublimation temperature of $1500~\kelvin$ at a column density
of $10^{21.63}~\cmmitwo$.  This is well below the ``Str\"omgren depth'' of $10^{21.23}~\cmmitwo$, where hydrogren is 50\%\ ionized.
The grain temperature as calculated by Cloudy would be 2000~\kelvin\ at this latter depth, and 3100~\kelvin\ at the cloud surface.
Reference to Figure 3 of \citet{ferland09} shows that in their model (similar to our Model 1), most of the optical 
\feii\ emission occurs below the sublimation depth of $10^{21.63}~\cmmitwo$, whereas most of the ultraviolet \feii\ emission
occurs at shallower depths.  Thus, evaporation of the grains occurs at a depth giving severe reduction of the optical 
but not the ultraviolet  \feii\  emission.  Dust will also not affect the emission from the highly ionized gas above the Str\"omgren depth,
such as \siiii\ and \siiv;  
and weak optical \feii\ could accompany a normal ratio of the ultraviolet \feii\ to \mgii\ lines.  The exact sublimation depth depends on grain size and composition, but the
qualitative pattern remains for a range of compositions and sizes included in the Cloudy output.  For different AGN with
different parameters, the sublimation point will occur at different depths in the ionization structure.   \citet{gaskell07} have also
discussed the affect of attenuation of the AGN continuum in the BLR on the radius of dust sublimation.

An alternative geometry for the BLR involves a radiatively driven wind from the accretion disk \citep{murray98}.  
In this picture, one issue is
whether grains can survive in the inflowing material to the radius where it is expelled in the wind.   Standard accretion disk physics
\citep{peterson97} gives for the disk effective temperature due to internal viscous dissipation
\begin{equation}
\teff = (10^{3.12}~\kelvin) (\mdot/\mdot_\mathrm{E})^{1/4} M_8^{-1/4} v_{3000}^{3/2},
\end{equation}
where $\mdot/\mdot_\mathrm{E}$ is the accretion rate relative to that giving the Eddington luminosity,  $M_8$ is the black hole mass
in units of $10^8~\msun$, and $v_{3000}$ is the orbital velocity at the radius of interest in units of 3000~\kms.   Thus, the
sublimation temperature for refractory grains  is reached at orbital velocities appropriate for the broad emission lines.  This may be consistent with differing degrees of grain evaporation in different objects.  (The mid-plane temperature of the disk will be higher.)
The above expression qualitatively agrees with the observed trend of stronger optical \feii\ with increasing Eddington ratio, but the dependence on
black hole mass may be problematic.   Moreover, it does not give such a natural $L^{0.5}$ dependence for the BLR radius as does the dust
sublimation model of \citet{netzer93}.   Note that in the disk-wind picture, the energy for the line emission still comes from photoionization by the central continuum; locally produced energy is insufficient at this shallow depth in the gravitational potential.  
Sublimation at an intermediate depth in the ionization structure, as discussed above, could also occur in the disk-wind model.

\subsection{Turbulent Velocity and Column Density}
\label{turb}

Local turbulence substantially affects the \feii\ spectrum in photoionization models by facilitating continuum and line-line fluorescence. 
Increasing the turbulence can increase the \feii\ strength and give better agreement between the predicted shape of the \feii\ blends and
observation \citep[and references therein]{baldwin04, bruhweiler08}.      \citet{bruhweiler08} find a factor of 2 increase in the UV \feii\ strength
relative to \mgii\ as the turbulence increases from 5 to 50~\kms (their Table 2).  When we decreased \vturb\ from 100~\kms\ to
zero in Model 1, the $\lambda 2500$ \feii\ band decreased a factor 10, and the $\lambda5000$ band decreased by a factor of 2.  If these
results are representative, they suggest that differing amounts of turbulence could make a substantial contribution to  the range of strength of \feii\
in AGN.  On the other hand, turbulence still appears to be inadequate to give the full observed range in optical \feii.  Moreover, \cite{baldwin04}
argue that substantial turbulence is required to fit the detailed shape of the UV \feii\ feature, so that there may be limited freedom to vary the turbulence.  A useful test may be to examine the observed shape of the UV and optical \feii\ blends as a function
of optical \feii\ strength, for comparison with photoionization models that vary either the turbulence or the gas-phase abundance of Fe.

The column density of the emitting clouds also has an important effect on the \feii\ emission.   \citet{ferland09} illustrate the increase
in \feii/\hbeta\ with increasing column density.  
They find that the minimum column density is $\sim10^{23}~\cmmitwo$ for gravity to overpower radiation pressure and
allow infall of clouds as found by  \citet{hu08}.    The UV \feii\ lines show little change above this column density, but the optical
\feii\ increases a factor $\sim0.5$~dex from  $\sim10^{23}~\cmmitwo$ to  $\sim10^{25}~\cmmitwo$.    Using arguments based on
virial determinations of the black hole mass in AGN, \citet{netzer08} also concludes that the column densities must 
substantially exceed $\sim10^{23}~\cmmitwo$ to avoid excessive effects of radiation pressure on the orbital velocities of the BLR clouds.
Thus, there may be limited freedom to vary the column density in order to produce the wide range of optical \feii\ strength observed.

The relative behavior of the optical and UV \feii\ bands may provide clues to the predominant cause of the range of \feii\ strength.
In our models, the UV and optical \feii\ both increase with Fe abundance, albeit more weakly for the UV blends.   However, increasing
the microturbulence increased the UV \feii\  by a greater factor that the optical \feii; and increasing the column density beyond
$\sim10^{23}~\cmmitwo$ mainly increases the optical \feii, as noted above.  \citet{shang07} give optical and \feii\ strengths for
a sample of AGN.  Their results show a much greater range in the EW of the optical \feii\ bands than in the UV bands, and a weak
anti-correlation between the optical and UV \citep[see also][]{wills85}.   The fact that the optical \feii\ shows a greater range
of intensity that the UV \feii\ may favor an explanation other than microturbulence.   One complication is whether the optical and UV lines originate at substantially different radii in the BLR, as suggested by some reverberation and line-width studies \citep[e.g.,][]{maoz93, hu08}.
\citet{ferland09} suggest that the observed optical \feii\ may be strongly affected by radiation escaping from the shielded
face of the photoionized clouds.  The observational and theoretical situation for \feii\ is complex, and further work will be needed to
devise definitive tests of the role of chemical abundances and physical conditions.

\section{Conclusion}
\label{sec:conclusion}

We have used composite SDSS spectra of AGN to examine the behavior of the narrow emission lines as a function of \feii\ strength.
The \fevii\ line shows only a weak increase with increasing \feii\ strength, indicating that the iron abundance contributes little to
the wide range of \feii\ strength in AGN. 
There is, however, a significant increase in the N/O abundance ratio with \feii\ strength, which suggests an increase in overall metallicity.
 There is little change in the level of ionization in the NLR as a function of \feii\ strength.  This, together with results of photoionization models,
 suggests that differences in the shape of the ionizing continuum, specifically the soft X-ray luminosity, are not the main drivers of the \feii\ strength.
 We propose that differences in the degree of depletion of Fe into grains in the low ionization portion of the BLR 
 are largely responsible for the weakness of
 \feii\ in some AGN, while it is strong in others.  This picture is consistent with the approximate coincidence of the BLR radius and the dust sublimation radius,
with indications that the BLR consists of material flowing inward from the dusty torus toward the central black hole, 
and with the variation of grain temperature with depth in the ionization structure of the BLR gas.

 The strength of \feii\ emission is a major component of the set of correlations known as ``Eigenvector 1'' (EV1) discussed by BG92.
 These include weak \oiii\ associated with strong \feii\ and narrower widths of the broad \hbeta\ line.  Radio loud AGN tend to have strong
 \oiii\ and weak \feii.   These trends have been the subject of many studies, but a good physical understanding of their origin remains lacking.
BG92 suggested that high column densities in the BLR enhance \feii\ while diminishing the ionizing radiation reaching the NLR.
 \citet{ludwig09}, in a spectral principal components analysis of AGN in SDSS, found the interpretation of the eigenvectors to be complicated.
 They argued that covering factor of the NLR was the likely cause of the range in \oiii\ strength.
 \citet{ferland09} suggest that the higher column densities required for infall in more luminous AGN can help to explain
 the correlation of \feii\ strength with $L/\led$.
 
 The interpretation of \feii\ strength in terms of dust depletion opens many questions for investigation.  Can support for this picture be found in the
 line intensities of other refractory elements?  Does the \feii\ emitting radius bear a different relationship to the
 dust sublimation radius for strong and weak \feii\ emitters?   What are the implications for the infrared emission of AGN?  
 What underlying causes lead to the correlations between \feii\ strength and other properties
 such as \oiii\ strength, radio emission, and Eddington ratio?   This paper does not offer answers to these
 larger questions, but the explanation of \feii\ strength in terms of gas-phase depletions gives a new context in which to address them.

\acknowledgments

We thank  Mark Botorff, Gary Ferland, Martin Gaskell, Richard Green,  Fred Hamann, Ari Laor, Hagai Netzer, and Bev Wills for helpful discussions and comments on the manuscript, and Alyx Stevens for assistance.
G.S. acknowledges support from the Jane and Roland Blumberg Centennial Professorship in Astronomy at the University of Texas at Austin.
We thank Karl Gebhardt for the composite spectrum program and Erin Bonning for the emission-line subtraction program used for the nitrogen line measurements.

Funding for the Sloan Digital Sky Survey (SDSS) has been provided by the Alfred P. Sloan Foundation, the Participating Institutions, the National Aeronautics and Space Administration, the National Science Foundation, the U.S. Department of Energy, the Japanese Monbukagakusho, and the Max Planck Society.  The SDSS is managed by the Astrophysical Research Consortium (ARC) for the Participating Institutions. The Participating Institutions are The University of Chicago, Fermilab, the Institute for Advanced Study, the Japan Participation Group, The Johns Hopkins University, the Korean Scientist Group, Los Alamos National Laboratory, the Max-Planck-Institute for Astronomy (MPIA), the Max-Planck-Institute for Astrophysics (MPA), New Mexico State University, University of Pittsburgh, University of Portsmouth, Princeton University, the United States Naval Observatory, and the University of Washington.


\clearpage

\clearpage

\clearpage

\clearpage

\clearpage


\begin{thebibliography}{38}
\expandafter\ifx\csname natexlab\endcsname\relax\def\natexlab#1{#1}\fi


\bibitem[{{Antonucci},{Miller}(1985)}] {antonuc85}
Antonucci, R. R. J., \& Miller, J. S. 1985, \apj, 297, 621

\bibitem[Baldwin et al.(1995)]{baldwin95} Baldwin, J., Ferland, G., Korista, K., \& Verner, D. 1995, \apjl, 455, L119

\bibitem[Baldwin et al.(2004)]{baldwin04} Baldwin, J. A., Ferland, G. J., Korista, K. T., Hamann, F., \& LaCluyz\'e, A. 2004, \apj, 615, 610

\bibitem[Bentz et al.(2009)]{bentz09} Bentz, M. C., Peterson, B. M., Netzer, H., Pogge, R. W., \& Vestergaard, M.  2009,
     \apj, 697, 160
 
\bibitem[Berrinton et al.(2000)]{berrington00} Berrington, K. A., Nakazaki, S., Norrington, P. H. 2000, \aaps,  142, 313
  
\bibitem[Boroson(2002)]{boroson02}  Boroson, T. A. 2002, \apj, 565, 78

\bibitem[Boroson \& Green(1992)]{boroson92}  Boroson, T. A. \& Green, R. F.  1992, \apjs, 80, 109 (BG92)

\bibitem[Bruhweiler \& Verner(2008)]{bruhweiler08}  Bruhweiler, F., \&  Verner, E. 2008, \apj, 675, 83

\bibitem[Collin \& Joly(2000)]{collin00} Collin, S., \& Joly, M. 2000, New Astronomy Reviews, 44, 531

\bibitem[Collin-Souffrin,  Hameury, \& Joly(1988)]{collin88} Collin, S., Hameury, J.-M., \& Joly, M. 1988, \astap, 205, 19

\bibitem[Delgado Inglada et al.(2009)]{delgado09} Delgado Inglada,  G., Rodr\'iquez, M., Mampaso, A., \& Vironen, K. 2009, \apj,     693, 1335

\bibitem[Ferguson et al.(1997a)]{ferguson97a} Ferguson, J. W., Korista, K. T., \& Ferland, G. J. 1997a, \apjs, 110, 287  

\bibitem[Ferguson et al.(1997b)]{ferguson97b} Ferguson, J. W., Korista, K. T., Baldwin, J. A., \& Ferland, G. J. 1997b, \apj, 487, 122

\bibitem[Ferland \& Persson(1989)]{ferland89} Ferland, G. J., \& Persson, S. E. 1989, \apj , 347, 656

\bibitem[Ferland et al.(1998)]{ferland98} Ferland, G. J., Korista, K.T., Verner, D.A., Ferguson, J.W., Kingdon, J.B. \&
Verner, E.M. 1998, PASP, 110, 761 

\bibitem[Ferland et al.(2009)]{ferland09} Ferland, G. J., Hu, C., Wang, J.-M., Baldwin, J. A., Porter, R. L., van Hoof, P. A. M., 
  \& Williams, R. J. R. 2009, \apjl, 707, L82

\bibitem[Garnett et al.(1995)]{garnett95} Garnett, D. R., Dufour, R. J., Peimbert, M., Torres-Peimbert, S., Shields, G. A., 
    Skillman, E. D., Terlevich, E.,   \& Terlevich, R. J. 1995, \apj, 449, L77

\bibitem[Gaskell(2009)]{gaskell09} Gaskell, M. 2009, New Astronomy Reviews, 53, 140
   
\bibitem[Gaskell et al.(2007)]{gaskell07}	Gaskell, C. M., Klimek, E. S., \& Nazarova, L. S. 2007, arXiv:0711.1025

\bibitem[Gaskell, Shields, \& Wampler(1981)]{gaskell81} Gaskell, C. M., Shields, G. A., \& Wampler 1981, \apj, 249, 443

\bibitem[Green(1998)]{green98} Green, P. J. 1998, \apj, 498, 170

\bibitem[Hamann \& Ferland(1993)]{hamann93} Hamann, F., \& Ferland, G. J. 1993, \apj, 418, 11

\bibitem[Hamann \& Ferland(1999)]{hamann99} Hamann, F., \& Ferland, G. 1999, \araa, 37, 487

\bibitem[Hamann et al.(2002)]{hamann02} Hamann, F., Korista, K. T., Ferland, G. J., Warner, C., \& Baldwin, J. 2002, \apj, 564, 592

\bibitem[Hamann et al.(2007)]{hamann07} Hamann, F., Warner, C., Dietrich, M., \& Ferland, G. 2007, in The Central Engine of Active Galactic Nuclei, ASP Conference Series, ed. Luis C. Ho and Jian-Min Wang, Vol. 373, p. 653

\bibitem[Hu et al.(2008)]{hu08} Hu, C., Wang, J.-M., Ho, L. C., Chen, Y.-M., Zhang, H.-T., Bian, W.-H., \& Xue, S.-J. 2008,  \apj, 687, 78

\bibitem[Kova\u{c}evi\'c et al.(2010)]{kovacevic10} Kova\u{c}evi\'c, J., Popovi\'c, L. C., \& Dimitrijevi\'c, M. S.  2010, \apjs, 189, 15

\bibitem[Kuehn et al.(2008)]{kuehn08}  Kuehn, C. A., Baldwin, J. A., Peterson, B. M., \& Korista, K. T. 2008, \apj, 673, 69

\bibitem[Kwan \& Krolik(1981)]{kwan81} Kwan, J., \& Krolik, J. H. 1981, \apj, 250, 478

\bibitem[Laor(2007)]{laor07}  Laor, A., in The Central Engine of Active Galactic Nuclei, ASP Conference Series, ed. Luis C. Ho and Jian-Min Wang, Vol. 373, p. 383

\bibitem[Laor \& Draine(1993)]{laor93}  Laor, A., \& Draine, B. T. 1993, \apj, 402, 441

\bibitem[Laor et al.(1997)]{laor97}  Laor, A., Fiore, F.,  Elvis, M., Wilkes, B. J., \& McDowell, J. C.1997, \apj, 477, 93

\bibitem[Lawrence et al.(1997)]{lawrence97} Lawrence, A., Elvis, M., Wilkes, B J., McHardy, I. \& Brandt, N. 1997, \mnras, 285, 879

\bibitem[Leighly et al.(2007)]{leighly07} Leighly, K. M., Halpern, J. P., Jenkins, E. B., \& Casebeer, D. 2007, \apjs, 173, 1

\bibitem[Ludwig et al.(2009)]{ludwig09}  Ludwig, R. R., Wills, B., Greene, J. E., \& Robinson, E. L. 2009, \apj, 706, 995

\bibitem[Maoz(1993)]{maoz93} Maoz, D., et al. 1993, \apj, 404, 576

\bibitem[Marziani et al.(2003)]{marziani03}  Marziani, P., Zamanov,  R. K., Sulentic, J. W., \& Calvani, M.  2003, \mnras, 435, 1133

\bibitem[Matsuoka, Kawara, \& Oyabu(2008)]{matsuoka08}  Matsuoka, Y., Kawara, K., \& Oyabu, S. 2008, \apj, 673, 62

\bibitem[Murray \& Chiang(1998)]{murray98} Murray, N., \& Chiang, J., 1998, \apj, 494, 125

\bibitem[Nagao et al.(2003)]{nagao03}  Nagao, T., Murayama, T., Shioya, Y., \& Taniguchi, Y., \aj, 125, 1729

\bibitem[Netzer(2008)]{netzer08}  Netzer, H., 2008, \apj, 695, 793

\bibitem[Netzer \& Laor(1993)]{netzer93}  Netzer, H., \& Laor, A. 1993, \apjl, 404, L51

\bibitem[Netzer \& Trakhtenbrot(2007)]{netzer07}  Netzer, H., \& Trakhtenbrot, B.  2007, \apj, 654, 754

\bibitem[Nussbaumer \& Osterbrock(1970)]{nussbaumer70} Nussbaumer, H., \& Osterbrock, D. E. 1970,  \apj, 161, 811

\bibitem[Osterbrock \& Ferland (2006)]{osterbr06} Osterbrock, D.~E., \& Ferland 2006, 
`Astrophysics of Gaseous Nebulae and Active Galactic Nuclei,'  2nd ed., Sausalito, California: University Science Books 

\bibitem[Peterson(1997)]{peterson97} Peterson, B. 1997, An Introduction to Active Galactic Nuclei (Cambridge: Cambridge Univ. Press)

\bibitem[{{Salviander} {et~al.}(2007){Salviander}, {Shields}, {Gebhardt}, \&
  {Bonning}}]{salviander07}
{Salviander}, S., {Shields}, G.~A., {Gebhardt}, K., \& {Bonning}, E.~W. 2007,
  \apj, 662, 131
  
\bibitem[Scott et al.(2004)]{scott04} Scott, J. E., Kriss, G. A., Brotherton, M., Green, R. F., Hutchings, J., Shull, J. M., 
   \& Wei, Z. 2004,  \apj, 615, 135

\bibitem[Shang et al.(2007)]{shang07} Shang, Z., Wills, B. J., Wills, D., \& Brotherton, M. S., 2007, \aj, 134, 294

\bibitem[Sigut \& Pradhan(2003)]{sigut03} Sigut, T. A. A., \& Pradhan, A. K. 2003, \apjs, 145, 15

\bibitem[Suganuma et al.(2006)]{suganuma06} Suganuma, M.,  et al. 2006, \apj, 639, 46

\bibitem[Sulentic, Marziani, \& Dultzin(2000)]{sulentic00} Sulentic, J. W., Marizani, P., \& Dultzin-Hacyan, D. 2000, \araa,  38, 521

\bibitem[van Zee et al.(1998)]{vanzee98} van Zee, L., Salzer, J. J., \& Haynes, M. P. 1998, \apjl, 497, L1

\bibitem[Vanden Berk et al.(2001)]{vandenberk01} Vanden Berk, D., et al. 2001
  \aj, 122, 549
  
\bibitem[Verner et al.(2003)]{verner03} Verner, E., Bruhweiler, F., Verner, D., Johansson, S., \& Gull, T. 2003, \apjl, 592, L59
 
\bibitem[Wills, Netzer, \& Wills(1985)]{wills85}  Wills, B. J., Netzer, H., \& Wills, D. 1985, \apj, 288, 94

 
\bibitem[Wills et al.(1999)]{wills99}  Wills, B. J., Laor, A.,  Brotherton, M. S., Wills, D., Wilkes, B. J., Ferland, G. J., 
  \& Shang, Z.  1999, \apjl, 515, L53


\end{thebibliography}
\end{document}